\documentclass[letterpaper]{article} % DO NOT CHANGE THIS
\usepackage{aaai2026}% DO NOT CHANGE THIS
\usepackage{times}  % DO NOT CHANGE THIS
\usepackage{helvet}  % DO NOT CHANGE THIS
\usepackage{courier}  % DO NOT CHANGE THIS
\usepackage[hyphens]{url}  % DO NOT CHANGE THIS
\usepackage{graphicx} % DO NOT CHANGE THIS
\usepackage{natbib}  % DO NOT CHANGE THIS AND DO NOT ADD ANY OPTIONS TO IT
\usepackage{caption} % DO NOT CHANGE THIS AND DO NOT ADD ANY OPTIONS TO IT
\usepackage{algorithm}
\usepackage{algorithmic}
\usepackage{booktabs}

\usepackage{newfloat}
\usepackage{listings}
\DeclareCaptionStyle{ruled}{labelfont=normalfont,labelsep=colon,strut=off} % DO NOT CHANGE THIS
\floatstyle{ruled}
\newfloat{listing}{tb}{lst}{}
\floatname{listing}{Listing}
\usepackage{enumitem}
\usepackage{multirow}
\usepackage{makecell}

\usepackage{todonotes}
\usepackage{subcaption}
\usepackage{xspace}

\newcommand{\historynetwork}{\ensuremath{\mathcal{G}}\xspace}

\begin{document}


\title{Gendered Pathways in AI Companionship: \\ Cross-Community Behavior and Toxicity Patterns on Reddit}
\author{
    %Authors
    % All authors must be in the same font size and format.
    Erica Coppolillo\textsuperscript{\rm 1, 2, 3} and 
    Emilio Ferrara\textsuperscript{\rm 1}
}
\affiliations{
    %Afiliations
    \textsuperscript{\rm 1}University of Southern California, Los Angeles, California\\
    \textsuperscript{\rm 2}University of Calabria, Rende, Italy\\
    \textsuperscript{\rm 3}ICAR-CNR, Rende, Italy\\
    % If you have multiple authors and multiple affiliations
    % use superscripts in text and roman font to identify them.
    % For example,
    % J. Scott Penberthy\textsuperscript{\rm 3},
    % George Ferguson\textsuperscript{\rm 4},
    % Hans Guesgen\textsuperscript{\rm 5}
    % Note that the comma should be placed after the superscript
    %erica.coppolillo@unical.it
%
% See more examples next
}

\nocopyright
\maketitle

\begin{abstract}
AI-companionship platforms are rapidly reshaping how people form emotional, romantic, and parasocial bonds with non-human agents, raising new questions about how these relationships intersect with gendered online behavior and exposure to harmful content. Focusing on the \textit{MyBoyfriendIsAI} (MBIA) subreddit, we reconstruct the Reddit activity histories of more than 3,000 highly engaged users over two years, yielding over 67,000 historical submissions. We then situate MBIA within a broader ecosystem by building a historical interaction network spanning more than 2,000 subreddits, which enables us to trace cross-community pathways and measure how toxicity and emotional expression vary across these trajectories. We find that MBIA users primarily traverse four surrounding community spheres (AI-companionship, porn-related, forum-like, and gaming) and that participation across the ecosystem exhibits a distinct gendered structure, with substantial engagement by female users. While toxicity is generally low across most pathways, we observe localized spikes concentrated in a small subset of AI-porn and gender-oriented communities. Nearly 16\% of users engage with gender-focused subreddits, and their trajectories display systematically different patterns of emotional expression and elevated toxicity, suggesting that a minority of gendered pathways may act as toxicity amplifiers within the broader AI-companionship ecosystem.
These results characterize the gendered structure of cross-community participation around AI companionship on Reddit and highlight where risks concentrate, informing measurement, moderation, and design practices for human-AI relationship platforms.
\end{abstract}

%\erica{What if we base the paper on the "loss" of the AI relationship?}

\section{Introduction}

Artificial–intelligence chatbots capable of sustaining intimate, emotional, and romantic interactions have grown dramatically in both visibility and adoption. Dedicated platforms such as Replika (\url{https://replika.com/}) and Character.ai (\url{https://character.ai/}) now support deeply personalized conversations that foster companionship, erotic engagement, and perceived emotional reciprocity~\cite{de2025ai, mlonyeni2025personal, chu2025illusionsintimacyemotionalattachment}. Prior work demonstrates that users often experience strong emotional attachment to these systems, rely on them for affective support, and sometimes consider them substitutes for human relationships \cite{emotional-ai, impacts-ai-companion}.  
Despite rising public and academic interest, most research on AI intimacy remains platform-specific, focusing on surveys or interviews with users of individual chatbot applications. As a result, we know little about how people who form such attachments behave across broader online ecosystems.

Theoretical frameworks on anthropomorphism, relational attachment, and parasocial bonds help explain why emotional connections to AI companions emerge \cite{Giles01082002, DERRICK2009352}. At the same time, work on algorithmic intimacy and adaptive conversational systems suggests that AI agents can influence users emotional states and expectations through sustained interaction \cite{hancock}. 

The \textit{MyBoyfriendIsAI} (MBIA) subreddit offers a compelling lens through which to examine these dynamics. Created in 2024, MBIA rapidly accumulated over $30,000$ followers, and tens of thousands of posts centered on emotional, romantic, and sexual experiences with AI partners. 
Although MBIA has raised recent interest~\cite{pataranutaporn2025myboyfriendaicomputational}, we are the first to explore this community at a larger scale, by examining its relationship to the wider constellation of online communities in which users participate. Specifically, in this work, we reconstruct two and a half years of Reddit activity (from January 2023 to September 2025) for more than $3,000$ highly active MBIA users. Using these data, we build a \textit{historical interaction network} of more than $2,000$ subreddits and $27,000$ directed edges, encoding the order in which users first posted across communities.
We organize our investigation around three main research questions:

\begin{itemize}[leftmargin=1cm]
    \item[\textbf{RQ1}:] What is the historical activity of MBIA users on Reddit? Which communities and pathways most frequently precede or follow engagement with AI-intimacy spaces?
    \item[\textbf{RQ2}:] How do these surrounding communities differ in terms of thematic focus, toxicity levels, and user gender composition?
    \item[\textbf{RQ3}:] Do MBIA users encounter gender-oriented or potentially radical/extremist communities along their trajectories, and how do their emotional and toxic behaviors differ within these spaces?
\end{itemize}

To answer these questions, we combine embedding-based topic modeling, toxicity estimation, and gender inference with large-scale network reconstruction. Our findings reveal that AI-romantic engagement is deeply embedded in a broader ecosystem of AI companionship, pornography, emotional support, gendered discourse, and gaming communities. Surprisingly, MBIA and much of its surrounding network are predominantly female, contradicting common narratives associating AI intimacy with male loneliness or incel-like subcultures \cite{massanari, LEOLIU2023107620}. 
While toxicity is generally low, localized spikes emerge in specific AI-porn and gender-oriented spaces, aligning with prior work on toxicity diffusion and exposure to harmful content \cite{toxicity, evolution-manosphere}. Gendered emotional patterns further suggest that users engage with these communities in meaningfully different ways.

Taken together, our work provides the first ecosystem-level, longitudinal perspective on AI romantic companionship on Reddit. We show how MBIA users behaviors reflect diverse online identities and pathways, highlighting both the societal implications and the emerging risks associated with human–AI emotional entanglements.

\section{Related Work}

% \begin{itemize}
%     \item \citet{pataranutaporn2025myboyfriendaicomputational}
%     \item \citet{chu2025illusionsintimacyemotionalattachment}: focuses on dedicated platforms such as Replika, Character.ai, and ChaiApp, thus targeting a specific, private LLM rather than a broader spectrum of models. 
%     \item \citet{quantifying-social-organization}
% \end{itemize}

\paragraph{Human–AI Companionship}
AI-driven companionship platforms such as Replika and Character.ai increasingly support intimate, romantic, and erotic interactions between humans and artificial agents. Prior research shows that users frequently report strong emotional attachment, comfort-seeking, and perceptions of reciprocal bonding with AI partners~\cite{skjuve2021my, PENTINA2023107600}. Studies further find that AI companions may influence wellbeing, loneliness, and expectations around human relationships~\cite{ta2020user}. Most prior work relies on platform-specific analyses or self-reported experiences, limiting insight into how AI intimacy fits into broader online ecosystems. Our study addresses this gap by examining longitudinal user behavior across thousands of subreddits, situating AI companionship in a broader network of online practices.

\paragraph{Algorithmic Intimacy}
Anthropomorphism and relational attachment theories, including parasocial relationships~\cite{Horton01081956} and social surrogacy, offer conceptual foundations for understanding emotional engagement with AI agents. Research on algorithmic intimacy and adaptive conversational systems highlights how AI-generated responses influence user behavior and attachment formation~\cite{skjuve2021my}. Our findings complement this line of work by showing how AI intimate interactions co-occur with participation in porn-oriented, emotional support, gender-hostile, and general forum communities, suggesting a contextualized and multi-layered form of AI-mediated intimacy.

\paragraph{Toxicity and Online Pathways}
Social computing research has extensively studied online toxicity~\cite{chandrasekharan2017yo, massanari} and user migration into extremist or hateful communities~\cite{mittos2020, ribeiro2021auditing}. Longitudinal patterns show that users often move along identifiable pathways when entering high-toxicity spaces~\cite{ribeiro2020evolution}. By reconstructing multi-year histories of MBIA users, our work extends these investigations to the domain of AI companionship, revealing that while toxicity is generally low, localized spikes emerge in specific AI-porn and gender-oriented communities.

\paragraph{Gender in Online Participation}
Gender substantially shapes online participation, emotional expression, and exposure to harm~\cite{massanari}. Prior studies show that subreddits often exhibit strong gender skews~\cite{ging2019alphas} and that gender-oriented communities can foster highly polarized discourse~\cite{ribeiro2020evolution}. Some recent work also compared extremist communities on Reddit, showing that no systematic discrepancies can be assessed between misogynistic and misandric ones~\cite{womenwhohate}. Our work contributes to this literature by uncovering a surprising predominance of female MBIA users, distinct emotional differences across genders in gender-oriented spaces, and highly polarized emotional patterns in specific subreddits.

\begin{figure}[!t]
    \centering
    \includegraphics[width=\linewidth]{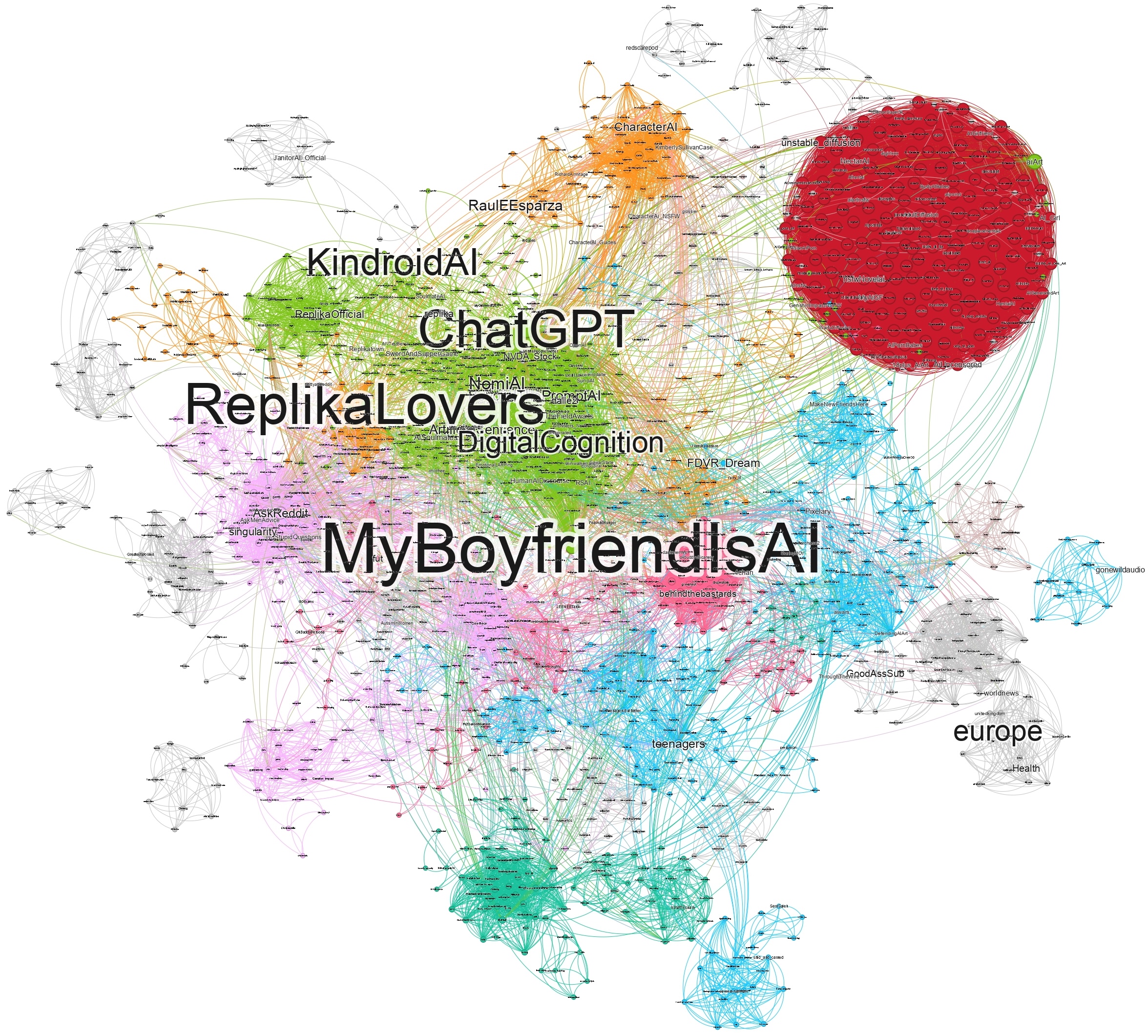}
    \caption{Historical network of the MBIA core users. Nodes are subreddits where users have posted in the considered timeframe. An edge exists between subreddit $v_1$ and subreddit $v_2$, if any user posts their first post on $v_1$ before their first post on $v_2$. Nodes are colored by modularity, with their size depending on degree. Label size reflects the posting activity within each subreddit (larger labels indicate a higher number of posts).}
    \label{fig:historical-network}
    \vspace{-.35cm}
\end{figure}

\section{Methodology}
In the following, we provide details on data acquisition and on the construction of the historical network of MBIA users.

\paragraph{Data} We retrieve the analyzed data from the public Reddit API. First, we collect over $5K$ submissions and $70K$ comments from the MBAI subreddit, posted between its creation in August 2024 and September 2025. To focus on the most engaged participants, we retain only users with more than five distinct interactions (submissions or comments), yielding a set of approximately $3K$ active users.
Starting from this core user base, we then gather all of their Reddit submissions dating back to January 2023, resulting in more than $67K$ posts. This allows us to reconstruct each user activity history over the preceding two and a half years, providing a comprehensive view of their platform behavior.

\paragraph{Historical Network} From the retrieved historical submissions, we construct a directed graph $\historynetwork = (V, E)$, where $V$ denotes the set of all subreddits appearing in the activity histories of the MBIA core users. We create a directed edge from $v_1$ to $v_2$ if a given user first post in $v_1$ occurred \textit{earlier} than their first post in $v_2$.
Semantically, $\historynetwork$ captures the navigational landscape of Reddit as traversed by MBIA users, representing the sequence in which these communities were first encountered throughout their historical activity. The resulting graph consists of $2,048$ nodes (subreddits) and $27,527$ edges (connections among communities). A visualization of the historical network is provided in Figure~\ref{fig:historical-network}.

%\paragraph{Settings} 

\begin{table}[!ht]
    \centering
    
    \caption{Subreddits in \historynetwork with the highest posting activity, along with their relative proportion across all the subreddits in the graph.}
    \label{tab:subreddit-activity}
    \begin{tabular}{lc}
    \toprule
    Subreddit &  Proportion \\
    \midrule
        ReplikaLovers & 0.034 \\
       ChatGPT & 0.03 \\
       KindroidAI & 0.02 \\
       DigitalCognition & 0.017\\
       europe & 0.017 \\
       NomiAI & 0.01 \\
       RaulEEsparza & 0.008 \\
       BeyondThePromptAI & 0.008 \\
       ArtificialSentience & 0.007 \\
    \bottomrule
    \end{tabular}
    \vspace{-.35cm}
\end{table}
\section{Results}

\begin{figure*}[!ht]
    \centering
    \begin{subfigure}[b]{0.49\textwidth} % [b] aligns subfigures at the bottom, 0.45\textwidth sets width
        \centering
        \includegraphics[width=\textwidth]{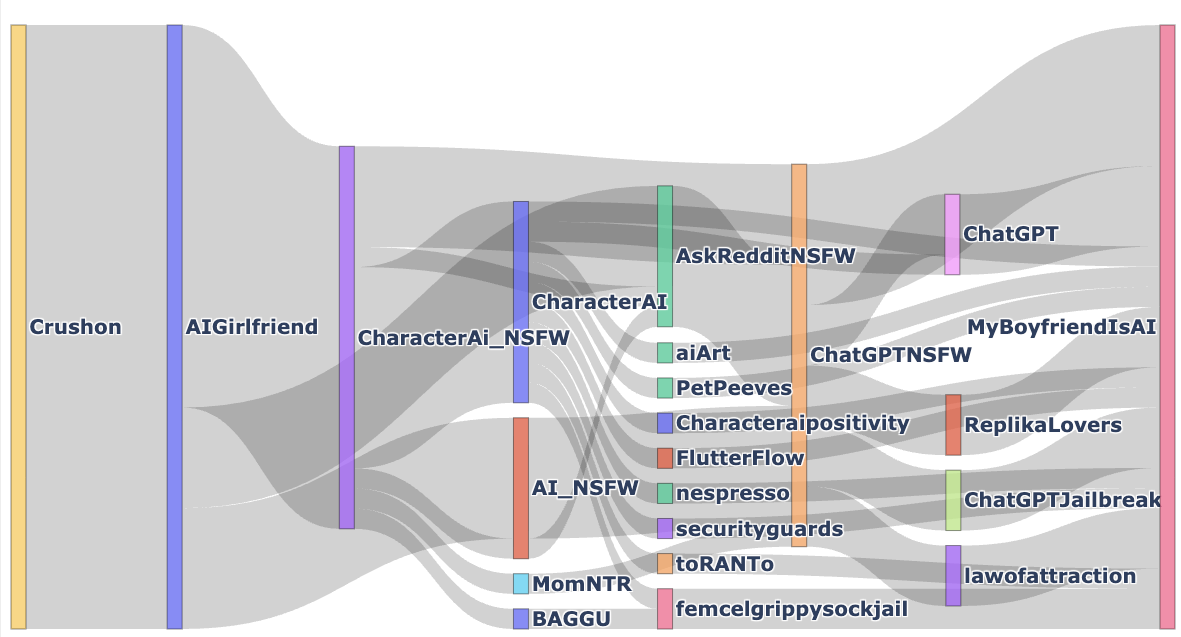} % image1.png is your image file
        \caption{}
        \label{fig:subfigA}
    \end{subfigure}
    %\hfill % Adds horizontal space between subfigures
    \begin{subfigure}[b]{0.49\textwidth}
        \centering
        \includegraphics[width=\textwidth]{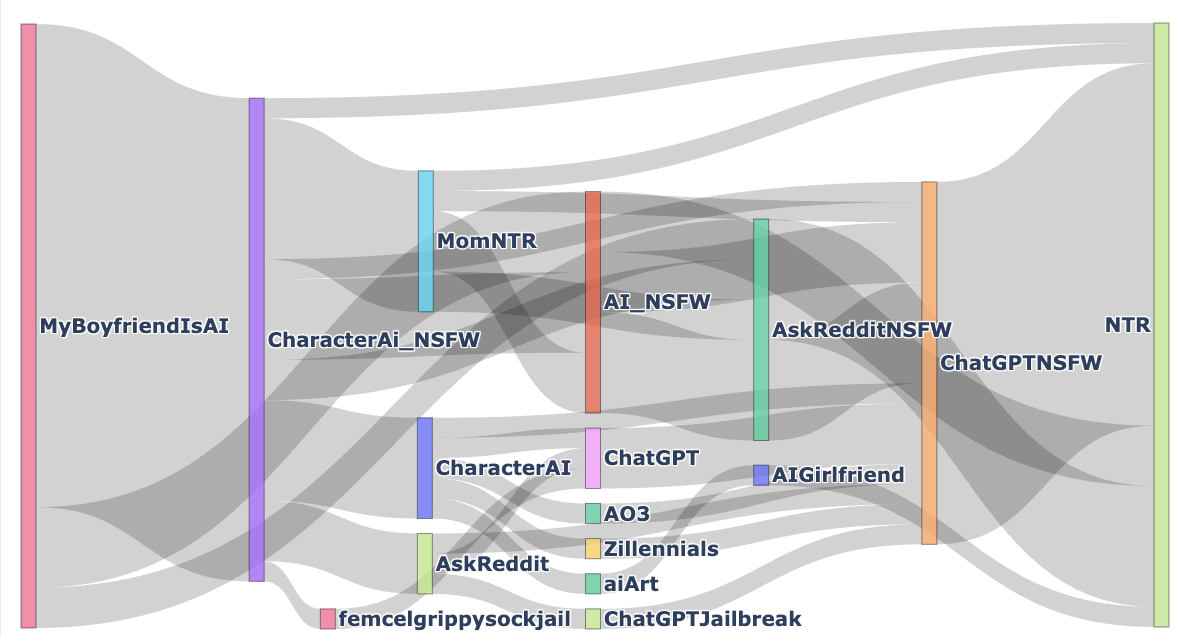} % image2.png is your image file
        \caption{}
        \label{fig:subfigB}
    \end{subfigure}
    \caption{Top-30 most common pathways in the historical network to and from MBIA, respectively.}
    \label{fig:common-pathways}
\end{figure*}

\paragraph{Unveiling MBIA Users Activity \rm{(RQ1)}} To investigate the historical activity of MBIA users on Reddit, we analyze the structure of the historical network \historynetwork, which captures the order in which users first engaged with different subreddits (Figure~\ref{fig:historical-network}). Table~\ref{tab:subreddit-activity} reports the top 10 subreddits in \historynetwork ranked by posting volume, together with their relative contribution to the total number of submissions in the network. We exclude MyBoyfriendIsAI from this ranking, as it would trivially appear as the most active subreddit by construction.
As expected, the subreddits exhibiting the highest activity levels predominantly revolve around human–chatbot companionship platforms (e.g., Replika, Kindroid) or AI-related discussions (e.g., ChatGPT), reflecting the thematic proximity of these communities to MBIA.

To gain deeper insight into the trajectory through which users arrive at, or depart from, MBIA, we examine the most frequent subreddit pathways encoded in the historical network. Specifically, we focus on the ego-component\footnote{Reference to be added.} of \historynetwork centered on MyBoyfriendIsAI, considering all nodes reachable within depth 5. This allows us to identify the dominant sequential patterns in users’ subreddit engagement leading to MBIA (Table~\ref{tab:mbia-target}) and originating from MBIA (Table~\ref{tab:mbia-source}). For each pathway, we report the subreddits involved, the thematic categories they represent, and their number of occurrences across user histories. To ensure semantic accuracy, subreddit topics were manually annotated based on their descriptions and by reviewing their most recent posts. We further provide in Figure~\ref{fig:common-pathways} the $30$ most common pathways leading to, and starting from MBIA, without filtering on the path lengths.

Interestingly, while most pathways are composed exclusively of AI-oriented or porn-oriented communities (an expected outcome given the nature of MBIA content) several notable exceptions emerge. A subset of pathways includes gender-focused or hateful communities such as ChatGPTJailbreak, Im21andDisillusioned, AskWomenOver30, femcelgrippysockjail, and 4bmovement. These cases highlight the diverse and sometimes problematic contexts in which MBIA users participate, pointing to a broader ecosystem of attitudes and interests beyond romantic AI interactions alone.

\begin{table*}[!ht]
        \centering
        \caption{The 5-Most popular subreddit paths in the submission graph, having ``MyBoyfriendIsAI'' as \textbf{target}/source node, respectively. Different path lengths are considered. For each path, we report the corresponding subreddit topics, the number of occurrences, and the proportion compared to all paths of the same length in the graph.}
        \label{tab:top-5-path-mbia-source}
        \begin{subtable}[t]{\linewidth}
        \resizebox{\linewidth}{!}{
        \begin{tabular}{cllcc}
            \toprule
            Length & Path & Topical Path & \#Occurrences & Proportion \\
            \midrule 
        \multirow{5}{*}{1}  & AIGirlfriend $\rightarrow$ MyBoyfriendIsAI & porn/relationship/ai $\rightarrow$ porn/relationship/ai & 21434 & 0.143 \\
  & aiArt $\rightarrow$ MyBoyfriendIsAI & art/ai $\rightarrow$ porn/relationship/ai & 11244 & 0.075 \\
  & ChatGPT $\rightarrow$ MyBoyfriendIsAI & ai/forum $\rightarrow$ porn/relationship/ai & 9974 & 0.067 \\
  & ArtificialSentience $\rightarrow$ MyBoyfriendIsAI & ai/forum $\rightarrow$ porn/relationship/ai & 9836 & 0.066 \\
  & ChatGPTPromptGenius $\rightarrow$ MyBoyfriendIsAI & ai/forum $\rightarrow$ porn/relationship/ai & 5874 & 0.039 \\
 \cmidrule(lr){1-5} 
\multirow{5}{*}{2}  & ChatGPT $\rightarrow$ ChatGPTPromptGenius $\rightarrow$ MyBoyfriendIsAI & ai/forum $\rightarrow$ ai/forum $\rightarrow$ porn/relationship/ai & 1184 & 0.008 \\
  & ChatGPT $\rightarrow$ ChatGPTJailbreak $\rightarrow$ MyBoyfriendIsAI & ai/forum $\rightarrow$ hateful/ai/forum $\rightarrow$ porn/relationship/ai & 1184 & 0.008 \\
  & ChatGPT $\rightarrow$ geminis $\rightarrow$ MyBoyfriendIsAI & ai/forum $\rightarrow$ forum $\rightarrow$ porn/relationship & 1182 & 0.008 \\
  & ChatGPT $\rightarrow$ diablo4 $\rightarrow$ MyBoyfriendIsAI & ai/forum $\rightarrow$ gaming $\rightarrow$ porn/relationship/ai & 1180 & 0.008 \\
  & ChatGPT $\rightarrow$ AlternativeSentience $\rightarrow$ MyBoyfriendIsAI & ai/forum $\rightarrow$ porn/relationship/ai $\rightarrow$ porn/relationship/ai & 1176 & 0.008 \\
 \cmidrule(lr){1-5} 
\multirow{5}{*}{3}  & ChatGPT $\rightarrow$ Im21andDisillusioned $\rightarrow$ ChatGPTPromptGenius $\rightarrow$ MyBoyfriendIsAI & porn/relationship/ai $\rightarrow$ incels/forum $\rightarrow$ ai/forum $\rightarrow$ porn/relationship/ai & 132 & 0.001 \\
  & ChatGPT $\rightarrow$ Murmuring $\rightarrow$ ChatGPTPromptGenius $\rightarrow$ MyBoyfriendIsAI & ai/forum $\rightarrow$ ai/forum $\rightarrow$ ai/forum $\rightarrow$ porn/relationship/ai & 132 & 0.001 \\
  & ChatGPT $\rightarrow$ thinkatives $\rightarrow$ ChatGPTPromptGenius $\rightarrow$ MyBoyfriendIsAI & ai/forum $\rightarrow$ forum/literature $\rightarrow$ ai/forum $\rightarrow$ porn/relationship/ai & 132 & 0.001 \\
  & ChatGPT $\rightarrow$ ChatGPTPro $\rightarrow$ ChatGPTPromptGenius $\rightarrow$ MyBoyfriendIsAI & ai/forum $\rightarrow$ ai/forum $\rightarrow$ ai/forum $\rightarrow$ porn/relationship/ai & 132 & 0.001 \\
  & ChatGPT $\rightarrow$ agi $\rightarrow$ ChatGPTPromptGenius $\rightarrow$ MyBoyfriendIsAI & ai/forum $\rightarrow$ ai/forum $\rightarrow$ ai/forum $\rightarrow$ porn/relationship/ai & 132 & 0.001 \\
 \cmidrule(lr){1-5} 
\multirow{5}{*}{4}  & ChatGPT $\rightarrow$ ChatGPTPromptGenius $\rightarrow$ ArtificialSentience $\rightarrow$ AlternativeSentience $\rightarrow$ MyBoyfriendIsAI & ai/forum $\rightarrow$ ai/forum $\rightarrow$ ai/forum $\rightarrow$ porn/relationship/ai $\rightarrow$ porn/relationship/ai & 14 & 0.0 \\
  & ChatGPT $\rightarrow$ ChatGPTPromptGenius $\rightarrow$ ArtificialSentience $\rightarrow$ aiArt $\rightarrow$ MyBoyfriendIsAI & ai/forum $\rightarrow$ ai/forum $\rightarrow$ ai/forum $\rightarrow$ art/ai $\rightarrow$ porn/relationship/ai & 14 & 0.0 \\
  & ChatGPT $\rightarrow$ ChatGPTPromptGenius $\rightarrow$ ArtificialSentience $\rightarrow$ ClaudeAI $\rightarrow$ MyBoyfriendIsAI & ai/forum $\rightarrow$ ai/forum $\rightarrow$ ai/forum $\rightarrow$ ai/forum $\rightarrow$ porn/relationship/ai & 14 & 0.0 \\
  & ChatGPT $\rightarrow$ ChatGPTPromptGenius $\rightarrow$ ArtificialSentience $\rightarrow$ OpenAI $\rightarrow$ MyBoyfriendIsAI & ai/forum $\rightarrow$ ai/forum $\rightarrow$ ai/forum $\rightarrow$ ai/forum $\rightarrow$ porn/relationship/ai & 14 & 0.0 \\
  & ChatGPT $\rightarrow$ ChatGPTPromptGenius $\rightarrow$ ArtificialSentience $\rightarrow$ technopaganism $\rightarrow$ MyBoyfriendIsAI & ai/forum $\rightarrow$ ai/forum  $\rightarrow$ ai/forum  $\rightarrow$ religion/relationship/ai $\rightarrow$ porn/relationship/ai & 14 & 0.0 \\
 \cmidrule(lr){1-5} 

        \end{tabular}}
        \caption{MBIA as target node}
        \label{tab:mbia-target}
        \end{subtable}
        \vfill
        \begin{subtable}[t]{\linewidth}
            \resizebox{\linewidth}{!}{
        \begin{tabular}{cllcc}
            \toprule
            Length & Path & Topical Path & \#Occurrences & Proportion \\
            \midrule 
        \multirow{5}{*}{1}  & MyBoyfriendIsAI $\rightarrow$ ChatGPT & porn/relationship/ai $\rightarrow$ ai/forum & 32296 & 0.263 \\
  & MyBoyfriendIsAI $\rightarrow$ ArtificialSentience & porn/relationship/ai $\rightarrow$ ai/forum & 17825 & 0.145 \\
  & MyBoyfriendIsAI $\rightarrow$ OpenAI & porn/relationship/ai $\rightarrow$ ai/forum & 15801 & 0.129 \\
  & MyBoyfriendIsAI $\rightarrow$ BeyondThePromptAI & porn/relationship/ai $\rightarrow$ porn/relationship/ai & 6328 & 0.052 \\
  & MyBoyfriendIsAI $\rightarrow$ RSAI & porn/relationship/ai $\rightarrow$ gaming & 5147 & 0.042 \\
 \cmidrule(lr){1-5} 
\multirow{5}{*}{2}  & MyBoyfriendIsAI $\rightarrow$ AskWomenOver30 $\rightarrow$ ChatGPT & porn/relationship/ai $\rightarrow$ women/relationship/forum $\rightarrow$ ai/forum & 2737 & 0.022 \\
  & MyBoyfriendIsAI $\rightarrow$ AIRelationships $\rightarrow$ ChatGPT & porn/relationship/ai $\rightarrow$ porn/relationship/ai $\rightarrow$ ai/forum & 2732 & 0.022 \\
  & MyBoyfriendIsAI $\rightarrow$ 4bmovement $\rightarrow$ ChatGPT & porn/relationship/ai $\rightarrow$ women/hateful $\rightarrow$ porn/relationship/ai & 2730 & 0.022 \\
  & MyBoyfriendIsAI $\rightarrow$ mbti $\rightarrow$ ChatGPT & porn/relationship/ai $\rightarrow$ gaming $\rightarrow$ ai/forum & 2730 & 0.022 \\
  & MyBoyfriendIsAI $\rightarrow$ ChatGPTNSFW $\rightarrow$ ChatGPT & porn/relationship/ai $\rightarrow$ porn/relationship/ai $\rightarrow$ ai/forum &  2730 & 0.022 \\
 \cmidrule(lr){1-5} 
\multirow{5}{*}{3}  & MyBoyfriendIsAI $\rightarrow$ CharacterAi\_NSFW $\rightarrow$ CharacterAI $\rightarrow$ ChatGPT & porn/relationship/ai $\rightarrow$ porn/relationship/ai $\rightarrow$ ai/forum $\rightarrow$ ai/forum & 215 & 0.002 \\
  & MyBoyfriendIsAI $\rightarrow$ CharacterAi\_NSFW $\rightarrow$ femcelgrippysockjail $\rightarrow$ ChatGPT & porn/relationship/ai $\rightarrow$ porn/relationship/ai $\rightarrow$ women/hateful $\rightarrow$ ai/forum & 215 & 0.002 \\
  & MyBoyfriendIsAI $\rightarrow$ ChatGPTPromptGenius $\rightarrow$ SaaS $\rightarrow$ ChatGPT & porn/relationship/ai $\rightarrow$ ai/forum $\rightarrow$ business/forum $\rightarrow$ ai/forum & 215 & 0.002 \\
  & MyBoyfriendIsAI $\rightarrow$ AskWomenOver30 $\rightarrow$ 4bmovement $\rightarrow$ ChatGPT & porn/relationship/ai $\rightarrow$ women/relationship/forum $\rightarrow$ women/hateful $\rightarrow$ ai/forum & 214 & 0.002 \\
  & MyBoyfriendIsAI $\rightarrow$ 4bmovement $\rightarrow$ LeopardsAteMyFace $\rightarrow$ ChatGPT & porn/relationship/ai $\rightarrow$ women/hateful $\rightarrow$ politics/hateful $\rightarrow$ ai/forum & 214 & 0.002 \\
 \cmidrule(lr){1-5} 
\multirow{5}{*}{4}  & MyBoyfriendIsAI $\rightarrow$ AI\_NSFW $\rightarrow$ AskRedditNSFW $\rightarrow$ ChatGPTNSFW $\rightarrow$ ChatGPT & porn/relationship/ai $\rightarrow$ porn/relationship/ai $\rightarrow$ porn/relationship/ai $\rightarrow$ porn/relationship/ai $\rightarrow$ ai/forum & 24 & 0.0 \\
  & MyBoyfriendIsAI $\rightarrow$ AI\_NSFW $\rightarrow$ ChatGPTNSFW $\rightarrow$ lawofattraction $\rightarrow$ ChatGPT & porn/relationship/ai $\rightarrow$ porn/relationship/ai $\rightarrow$ porn/relationship/ai $\rightarrow$ relationship/forum $\rightarrow$ ai/forum & 24 & 0.0 \\
  & MyBoyfriendIsAI $\rightarrow$ AskRedditNSFW $\rightarrow$ ChatGPTNSFW $\rightarrow$ lawofattraction $\rightarrow$ ChatGPT & porn/relationship/ai $\rightarrow$ porn/relationship/ai $\rightarrow$ porn/relationship/ai $\rightarrow$ relationship/ai $\rightarrow$ ai/forum & 24 & 0.0 \\
  & MyBoyfriendIsAI $\rightarrow$ CharacterAi\_NSFW $\rightarrow$ AI\_NSFW $\rightarrow$ ChatGPTNSFW $\rightarrow$ ChatGPT & porn/relationship/ai $\rightarrow$ porn/relationship/ai $\rightarrow$ porn/relationship/ai $\rightarrow$ porn/relationship/ai $\rightarrow$ ai/forum & 24 & 0.0 \\
  & MyBoyfriendIsAI $\rightarrow$ CharacterAi\_NSFW $\rightarrow$ AskRedditNSFW $\rightarrow$ ChatGPTNSFW $\rightarrow$ ChatGPT & porn/relationship/ai$\rightarrow$ porn/relationship/ai $\rightarrow$ porn/relationship/ai $\rightarrow$ porn/relationship/ai $\rightarrow$ ai/forum & 24 & 0.0 \\
 \cmidrule(lr){1-5} 

        \end{tabular}
        }
        \caption{MBIA as source node}
        \label{tab:mbia-source}
        \end{subtable}
\end{table*}
        %\vspace{-.35cm}

This analysis motivates our subsequent investigations, which aim to further characterize the subreddits present in the network, in terms of their topical domains, toxicity levels, and the gender distribution of their users.

\begin{figure}[!ht]
    \centering
    \includegraphics[width=0.8\columnwidth]{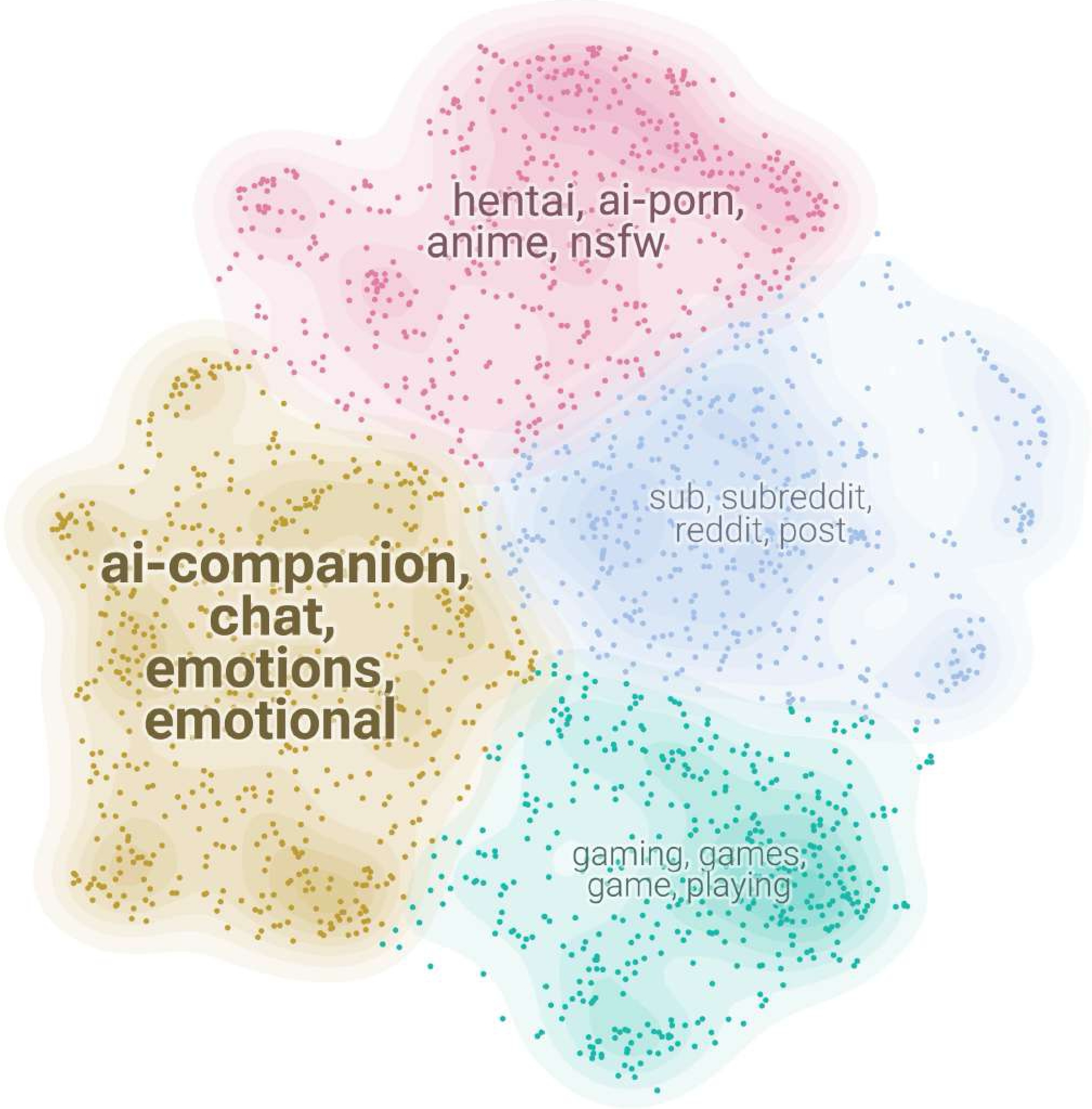}
    \caption{Topic map of the users historical subreddits, computed via BERTopic. For each subreddit, we consider its name and description, or $10$ randomly sampled posts when the description was not available. Embeddings are generated via all-MiniLM-L6-v2, while clusters are computed via KMeans. Elbow method has been used to find the optimum number of clusters.}
    \label{fig:topic-map}
\end{figure}

\paragraph{Characterizing Network Subreddits \rm{(RQ2)}} We further examine the subreddits in \historynetwork through the lens of their general topics. To do so, we analyze both the subreddit names and their public descriptions; when a description was unavailable, we additionally sampled 10 random posts to infer the subreddit semantic content.

\begin{figure}
    \centering
    \includegraphics[width=\linewidth]{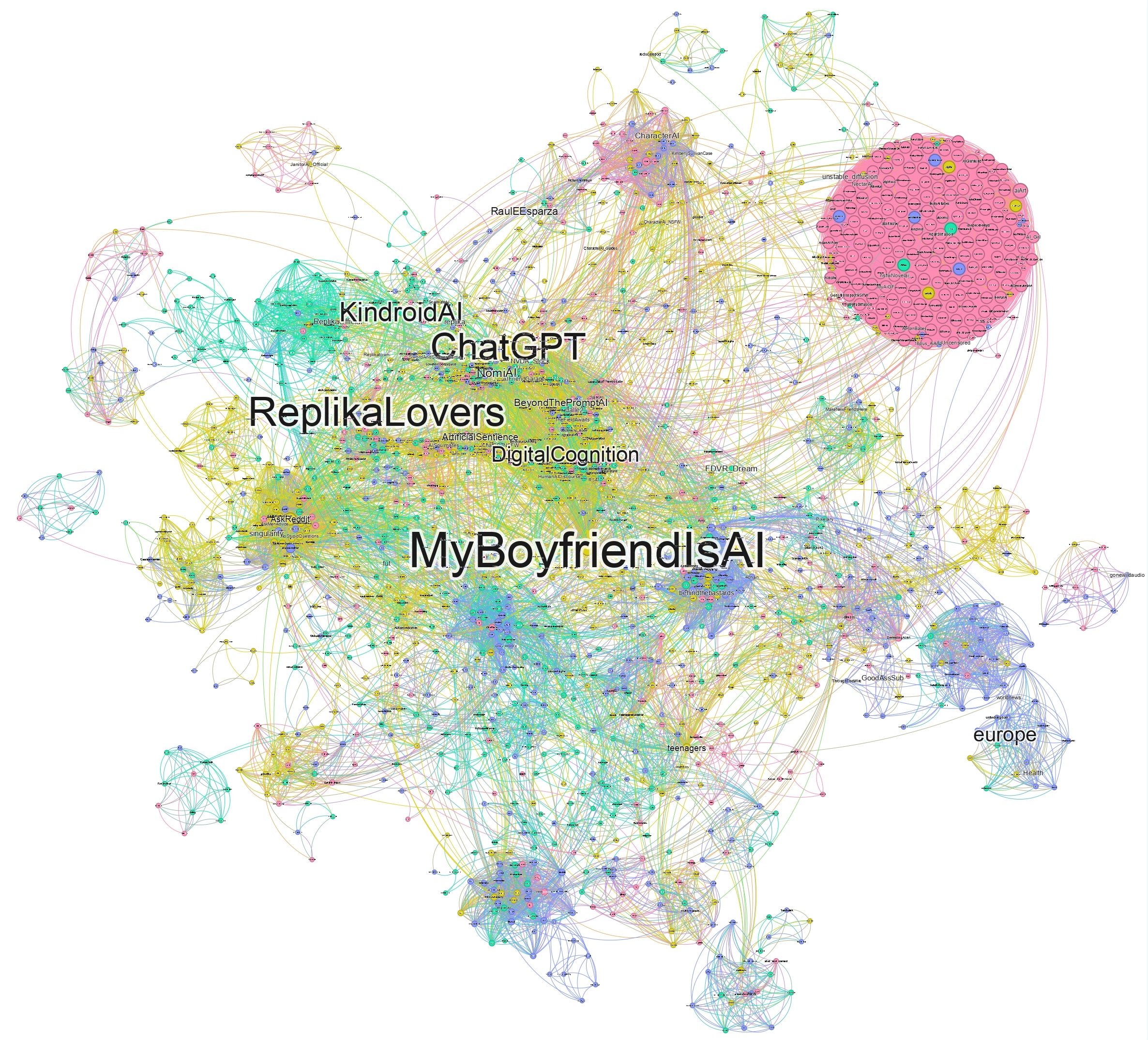}
    \includegraphics[width=\linewidth]{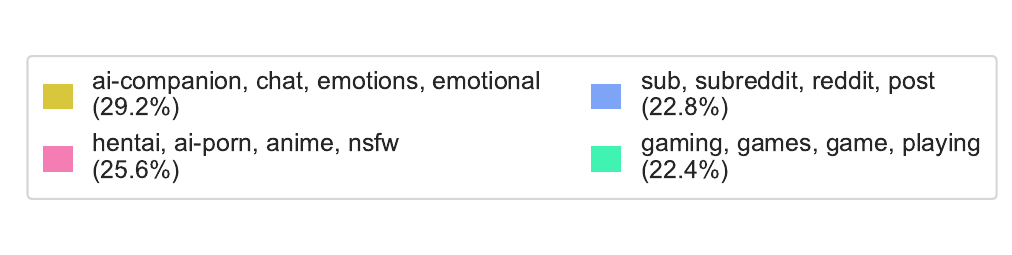}
    \caption{History network where nodes are colored by topic.}
    \label{fig:historical-network-topics}
\end{figure}

Similarly to~\cite{pataranutaporn2025myboyfriendaicomputational}, we compute sentence embeddings using the transformer-based model \texttt{all-MiniLM-L6-v2}\footnote{\url{https://huggingface.co/sentence-transformers/all-MiniLM-L6-v2}}~\cite{wang2020minilmdeepselfattentiondistillation}, and cluster these embeddings via KMeans~\cite{Jin2010}. The optimal number of clusters is determined using the Elbow method~\cite{elbow}. We further apply BERTopic~\cite{grootendorst2022bertopicneuraltopicmodeling} to extract the most representative topics associated with each cluster. The resulting topical map is shown in Figure~\ref{fig:topic-map}, while the representation of \historynetwork colored by topic is provided in Figure~\ref{fig:historical-network-topics}. Four dominant thematic communities emerge: AI companionship (29.2\% subreddits), porn-related content (25.6\%), forum-style discussions (22.8\%), and gaming (22.4\%). Notably, the porn-oriented community represent the \textit{densest} connected component in the graph, with more than $50\%$ of edges involved (top-right in the figure). 

Next, we assess the average toxicity of the historical network subreddits. For this analysis, we employ \texttt{detoxify}\footnote{\url{https://pypi.org/project/detoxify/}}~\cite{Detoxify}, a BERT-based model trained for toxic comment classification and exhibiting a considerable accuracy ($\sim~0.99$ mean AUC score). Given an input text, the model outputs a toxicity score in $[0,1]$, indicating the degree of toxicity expressed.

We apply this classifier to all posts contained in each subreddit of \historynetwork and compute the average toxicity per subreddit.
We additionally compute the relative change in toxicity when moving from one subreddit to another. Positive (resp. negative) values indicate that the destination subreddit is more (resp. less) toxic than the source.

\begin{figure}[!ht]
    \centering
    \includegraphics[width=\linewidth]{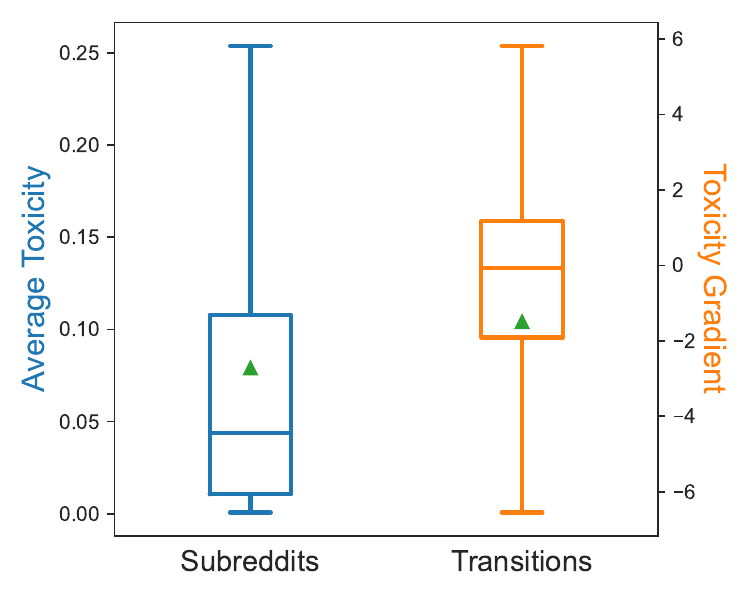}
    \caption{Toxicity distribution of the average toxicity of the subreddits and in terms of relative change across subreddits. Outliers were removed for readability.}
    \label{fig:toxicity-nodes-edges}
\end{figure}

Figure~\ref{fig:toxicity-nodes-edges} displays the toxicity distribution computed over the nodes (subreddits) and across the edges (transitions) of \historynetwork. The respective means and standard deviations are $\mu = 0.079, \sigma = 0.1$ on the subreddits, and $\mu = -1.5, \sigma = 29.03$ on the transitions. In the figure, we removed the outliers to ensure readability. 

    \begin{table}[!ht]
        \centering
        \caption{Top-$10$ most toxic subreddits in MBIA users history network. For each subreddit, we report the corresponding average toxicity score and the length of the shortest path to/from ``MyBoyfriendIsAI''.}
        \label{tab:top-10-toxic-subreddits}
        \resizebox{.8\linewidth}{!}{
        \begin{tabular}{lccc}
            \toprule
            Subreddit & Toxicity &  Shortest Path Length \\ 

            \midrule
        gaycheaters & 0.88 &  3 \\ 
DaisyRidleyLust2 & 0.82 &  1 \\ 
chaoticgood & 0.75 &  3 \\ 
RoleplayFunLimitless & 0.74 &  4 \\ 
MomNTR & 0.69 &  1 \\ 
IncelTear & 0.61 &  3 \\ 
Roleplayheaven & 0.6 &  4 \\ 
neighborsfromhell & 0.57 &  1 \\ 
trulynorules & 0.55 &  2 \\ 
SluttyConfessions & 0.54 &  3 \\ 
%ControversialOpinions & 0.52 &  3 \\ 

    \bottomrule
    \end{tabular}}
    \end{table}

Overall, most subreddits exhibit low toxicity levels, and transitions between them are generally smooth, showing no sharp variations when users move from one subreddit to another.
Despite this overall benign landscape, we identify a subset of subreddits with alarmingly high toxicity, considerably above the mean score of 0.08. Table~\ref{tab:top-10-toxic-subreddits} reports the top 10 most toxic subreddits, along with their average toxicity and shortest-path distance from MyBoyfriendIsAI. Notably, despite most of the subreddits focus on sexual content, one is a gender-oriented community (IncelTear).

\begin{figure}[!ht]
\centering
\includegraphics[width=\linewidth]{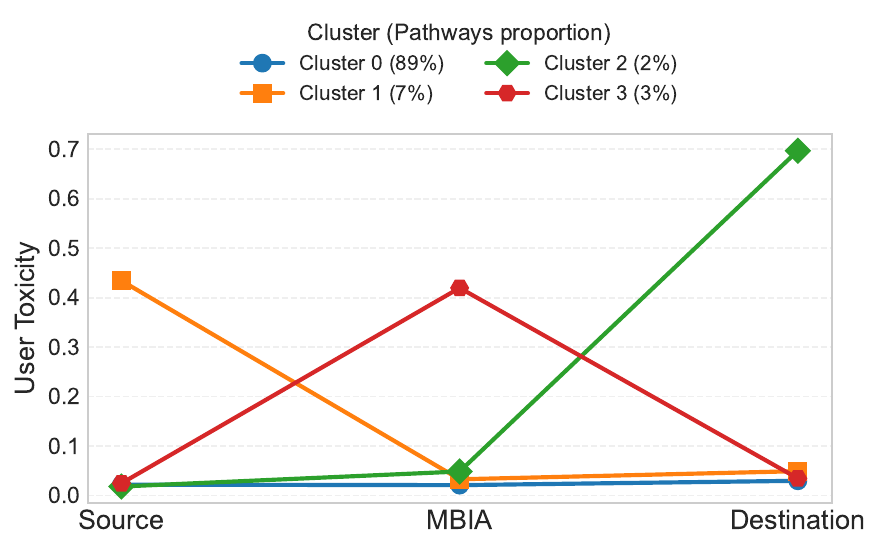}
\caption{User-level toxicity trajectories computed over the 3-length pathways traversing MBIA.}
\label{fig:user-toxicity-centered-mbia}
\end{figure}

Given that several of these communities lie close to MBIA, we conduct a more fine-grained analysis to determine whether they may function as ``toxicity gateways'' for MBIA users. To this end, we examine all subreddit pathways of length 2 leading to or departing from MyBoyfriendIsAI, and we compute the user-level average toxicity trajectory along each pathway. These trajectories are then clustered using KMeans, with the Elbow method identifying four optimal clusters. Figure~\ref{fig:user-toxicity-centered-mbia} visualizes the resulting clusters.

The results indicate that the vast majority of pathways (89\%) remain consistently low in toxicity. Notably, toxicity spikes occur in a minority of cases: 7\% of trajectories show elevated toxicity \textit{before} reaching MBIA; 2\% show increased toxicity \textit{after} MBIA; and 3\% exhibit a toxicity peak at MBIA itself.

To better understand these dynamics, we isolate the pathways with the highest and lowest relative toxicity gaps (Tables~\ref{tab:table_a} and~\ref{tab:table_b}). This analysis reveals that the most toxic sources and destinations predominantly correspond to AI-porn–oriented communities, reinforcing the observation that toxicity, when present, is concentrated within a narrow subset of thematically extreme subreddits.

\begin{table*}[!ht]
        \centering
        \caption{Top-10 trajectories traversing MBIA sorted by user toxicity relative gap. Positive (resp. negative) values indicate a toxicity increase (resp. decrease) registered by the user moving from ``Source'' to ``Destination''.}
        \label{tab:placeholder}
        \begin{subtable}[t]{0.48\linewidth}

        \resizebox{\linewidth}{!}{
        \begin{tabular}{@{}ll@{}ccc@{}}
            \toprule 
            Source & Destination & Source User Toxicity & Destination User Toxicity & Relative Gap \\

            \midrule

        Chatbots & MomNTR & 0.001 & 0.692 & 1166.427\\ 
LocalLLaMA & ClaudeAI & 0.001 & 0.783 & 1006.636\\ 
femcelgrippysockjail & OpenAI & 0.001 & 0.493 & 594.691\\ 
Chatbots & AskRedditNSFW & 0.001 & 0.345 & 582.211\\ 
ArtificialSentience & ClaudeAI & 0.002 & 0.783 & 365.729\\ 
Chatbots & NTR & 0.001 & 0.215 & 361.146\\ 
chaosmagick & fictobots & 0.001 & 0.254 & 297.041\\ 
ChatGPTPromptGenius & ClaudeAI & 0.003 & 0.783 & 294.242\\ 
ChatGPT & ChatGPTNSFW & 0.001 & 0.332 & 267.228\\ 
AskReddit & 4bmovement & 0.001 & 0.231 & 238.221\\ 
\bottomrule
        \end{tabular}
        }
        \caption{Highest user toxicity relative gap.}
        \label{tab:table_a}
        \end{subtable}
        \hfill %{\fill}
\begin{subtable}[t]{0.48\linewidth}
%\flushright
\resizebox{\linewidth}{!}{
        \begin{tabular}{@{}ll@{}ccc@{}}
            \toprule 
            Source & Destination & Source User Toxicity & Destination User Toxicity & Relative Gap \\

            \midrule

        AskReddit & adhdwomen & 0.573 & 0.001 & -518.776\\ 
ChatGPTNSFW & AISoulmates & 0.839 & 0.003 & -240.455\\ 
AMA & okbuddygganbu & 0.167 & 0.001 & -206.829\\ 
AskReddit & ChatGPTPromptGenius & 0.069 & 0.001 & -93.236\\ 
ChatGPT & ChatGPTNSFW & 0.167 & 0.002 & -88.104\\ 
diablo4 & ChatGPT & 0.032 & 0.001 & -51.989\\ 
ChatGPT & OpenAI & 0.041 & 0.002 & -22.016\\ 
ChatGPT & AISoulmates & 0.019 & 0.001 & -16.682\\ 
ChatGPT & ArtificialSentience & 0.02 & 0.001 & -16.204\\ 
replika & AIFriendGarage & 0.019 & 0.001 & -13.872\\ 
\bottomrule
        \end{tabular}
        }
        \caption{Lowest user toxicity relative gap.}
        \label{tab:table_b}
        \end{subtable}
        \end{table*}

Finally, we characterize the historical network in terms of user \textit{gender}. Because Reddit does not provide demographic information, we rely on a DeBERTa-based text classifier \cite{coban_gender_prediction_model_from_text_2025}
 that infers gender from linguistic cues. We apply this model to all submissions authored by each user and assign a gender label based on the most frequently predicted category across their posts.
Surprisingly, the results indicate that the majority of the MBIA core user base is female ($\sim$63\%).

We then propagate gender information to the network level by tagging each node (subreddit) in \historynetwork with the most prevalent gender among its active users.

To assess the robustness of these gender labels, we compare them with the subreddit-level gender annotations released by~\cite{quantifying-social-organization}, which provide gender distributions for over 10,000 subreddits. We observe a 67\% agreement between our predicted labels and those reported in the dataset. For subreddits where disagreement occurs and external annotations are available, we adopt the labels from~\cite{quantifying-social-organization}. For subreddits absent from their dataset, we retain our classifier-based assignments.

\begin{figure}[!ht]
    \centering
    \includegraphics[width=\linewidth]{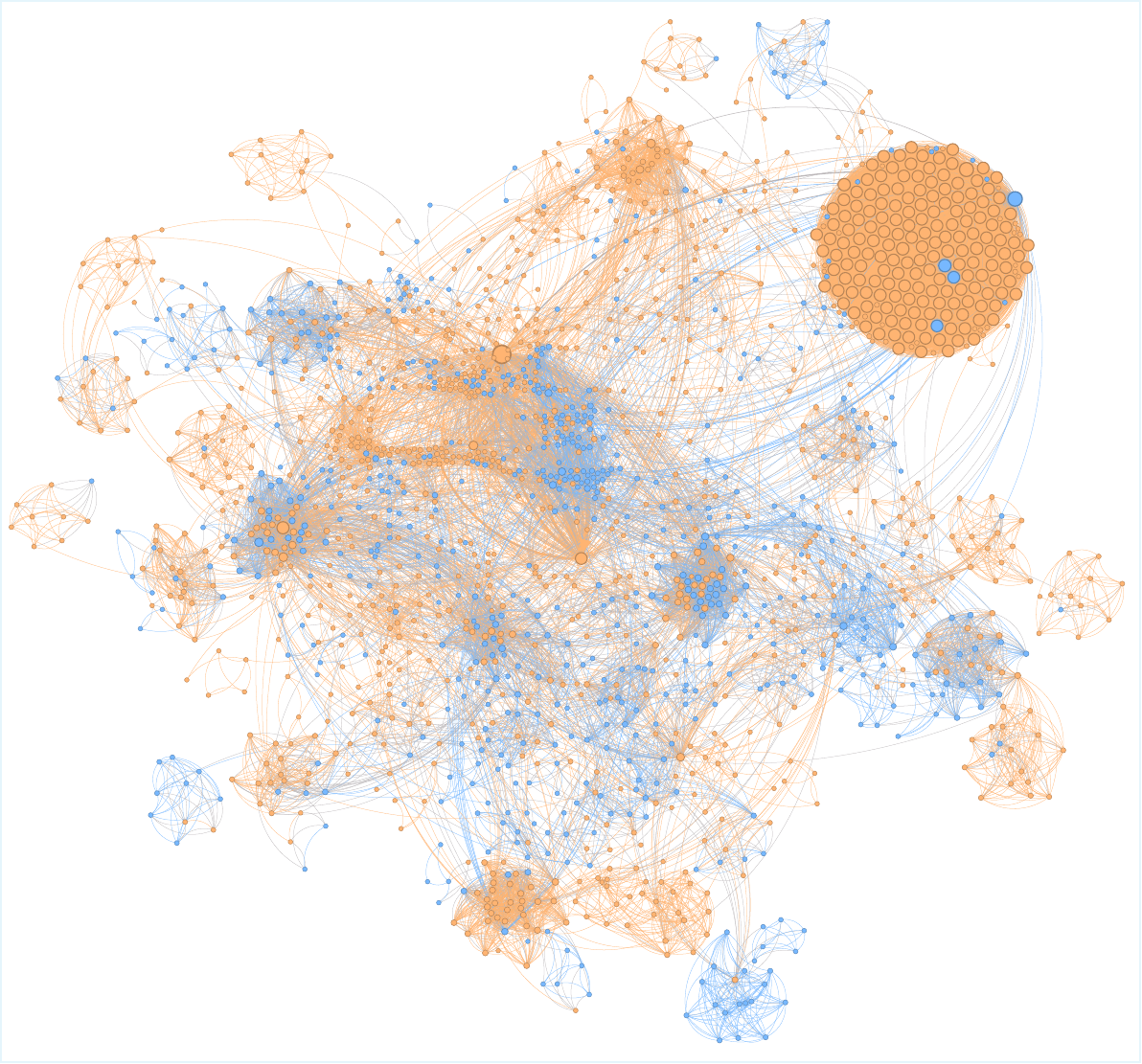}
    \includegraphics[width=0.5\linewidth]{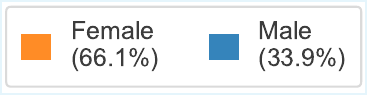}
    \caption{History network where nodes are colored according to the majority user base gender.}
    \label{fig:historical-network-gender}
\end{figure}

Consistent with our user-level findings, \historynetwork emerges as predominantly female, with approximately 66\% of subreddits showing a higher proportion of female active users. Figure~\ref{fig:historical-network-gender} visualizes \historynetwork colored by majority-gender.
Interestingly, even the porn-oriented cluster (top-right of the figure) is largely female-dominated, highlighting an unexpected demographic pattern within this thematic community. 

\paragraph{Exploring Gender-oriented Communities \rm{(RQ3)}} Our final analysis focuses on identifying and examining the subreddits within \historynetwork that are explicitly \textit{gender-oriented}, that is, communities centered around male- or female-related themes. To detect them, we construct a list of gender-related keywords (e.g., ``men'', ``fem'', ``girl'') and apply this lexicon to the subreddit names in \historynetwork. We then manually review the results to filter out subreddits whose names contain relevant keywords but do not primarily focus on gender (e.g., communities celebrating historical figures).

\begin{table}[!ht]
    \centering
    
    \caption{Gender-oriented subreddits present in the submission history of MBIA users. For each subreddit, we report the number of posts/comments, the number of distinct MBIA users, and whether it is known or self-declared to be radical/extremist.}
    \label{tab:gender-subreddits}
    \resizebox{\columnwidth}{!}{
    \begin{tabular}{@{}l@{}ccc@{}}
    \toprule
    Subreddit & \#Posts/Comments & \#MBIA Users (\%) & Radical \\
    \midrule
        4bmovement & 499 & 30 (1.0) & Yes \\
        AskMen & 3,012 & 149 (5.0) & No \\ 
        AskMenAdvice & 4,318 & 185 (6.2) & No \\
        AskMenOver30 & 502 & 62 (2.1) & No\\
        AskMenRelationships & 114 & 4 (0.1) & No\\
        AskWomenNoCensor & 985 & 29 (1.0) & No\\
        AskWomenOver30 & 2,440 & 62 (2.1) & No\\
        Feminism & 260 & 44 (1.5) & Yes\\
        ForeverAloneWomen & 679 & 29 (1.0) & No\\
        IncelTear & 365 & 15 (0.5) & No\\
        IncelTears & 259 & 30 (1.0) & No\\
        MensRights & 481 & 25 (0.8) & Yes\\
        TheGirlSurvivalGuide & 598 & 44 (1.5) & No\\
        WomenAreNotIntoMen & 321 & 7 (0.2) & No\\
        femcelgrippysockjail & 1,024 & 45 (1.5) & Yes\\
        women & 823 & 38 (1.3) & No\\
    \midrule
    Total & 16,680 & 475 (15.8) & - \\
    \bottomrule
    \end{tabular}
    }
\end{table}

% Prop of mbia users in subreddit 4bmovement: 30 (1.0 %
% #Posts/comments: 499
% ***
% Prop of mbia users in subreddit AskMen: 149 (5.0 %
% #Posts/comments: 3012
% ***
% Prop of mbia users in subreddit AskMenAdvice: 185 (6.0 %
% #Posts/comments: 4318
% ***
% Prop of mbia users in subreddit AskMenOver30: 62 (2.0 %
% #Posts/comments: 502
% ***
% Prop of mbia users in subreddit AskMenRelationships: 4 (0.0 %
% #Posts/comments: 114
% ***
% Prop of mbia users in subreddit AskWomenNoCensor: 29 (1.0 %
% #Posts/comments: 985
% ***
% Prop of mbia users in subreddit AskWomenOver30: 62 (2.0 %
% #Posts/comments: 2440
% ***
% Prop of mbia users in subreddit Feminism: 44 (1.0 %
% #Posts/comments: 260
% ***
% Prop of mbia users in subreddit ForeverAloneWomen: 29 (1.0 %
% #Posts/comments: 679
% ***
% Prop of mbia users in subreddit IncelTear: 15 (0.0 %
% #Posts/comments: 365
% ***
% Prop of mbia users in subreddit IncelTears: 30 (1.0 %
% #Posts/comments: 259
% ***
% Prop of mbia users in subreddit MensRights: 25 (1.0 %
% #Posts/comments: 481
% ***
% Prop of mbia users in subreddit TheGirlSurvivalGuide: 44 (1.0 %
% #Posts/comments: 598
% ***
% Prop of mbia users in subreddit WomenAreNotIntoMen: 7 (0.0 %
% #Posts/comments: 321
% ***
% Prop of mbia users in subreddit femcelgrippysockjail: 45 (1.0 %
% #Posts/comments: 1024
% ***
% Prop of mbia users in subreddit women: 38 (1.0 %
% #Posts/comments: 823
% ***

This procedure yields 16 gender-oriented communities, reported in Table~\ref{tab:gender-subreddits}. For each subreddit, we list the number of posts/comments produced by MBIA users, the percentage of MBIA core users active within it, and whether the community is known to be, or self-describes as, radical or extremist. Among the 16 identified subreddits, four fall into this category: 4bmovement\footnote{\url{https://en.wikipedia.org/wiki/4B_movement}}, inspired by a radical Korean feminist group~\cite{yoon2022beneath}; Feminism\footnote{\url{https://en.wikipedia.org/wiki/R/Feminism}}, a feminist political subreddit discussing women's issues; MensRights\footnote{\url{https://en.wikipedia.org/wiki/Controversial_Reddit_communities}, a self-declared ``antifeminist'' community}; and femcelgrippysockjail, self-described as dedicated to ``evil women''.
Overall, nearly 16\% of MBIA users participate in at least one of these gender-oriented communities.

\begin{figure}[!ht]
\centering
\includegraphics[width=\linewidth]{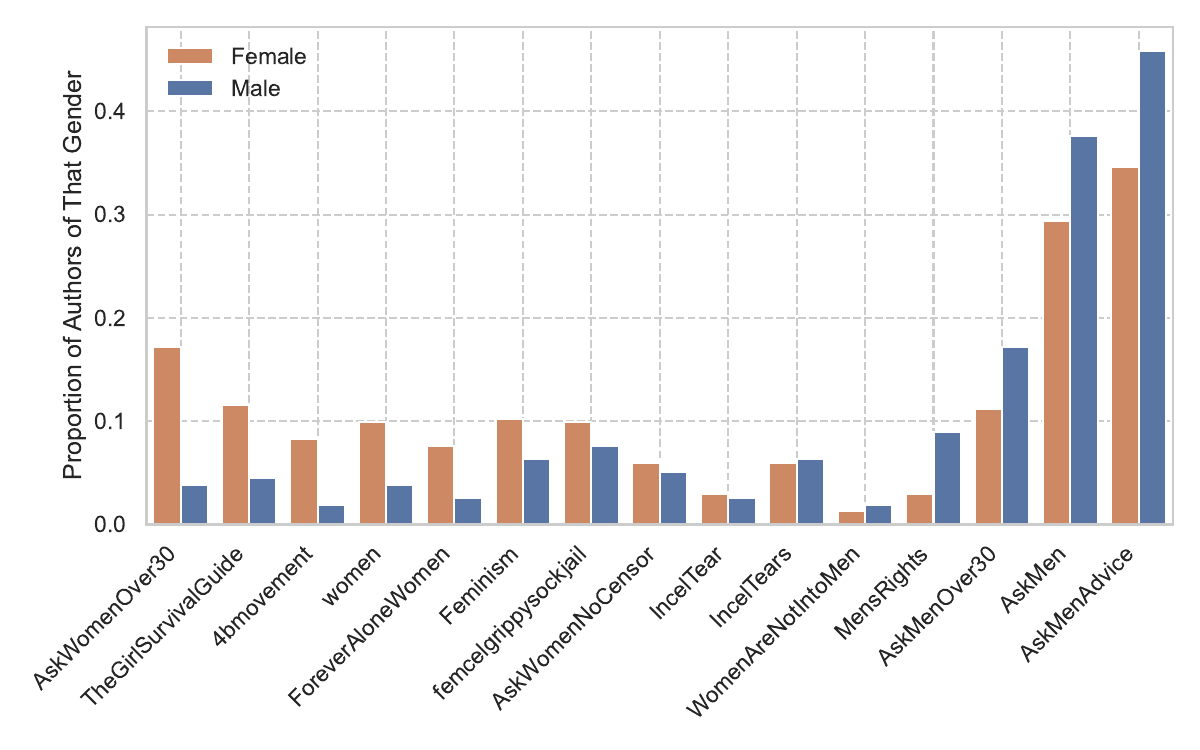}
\caption{Proportion of female/male users per gender-oriented community. Proportions sum to 1 per subreddit.}
\label{fig:gender-distributions-communities}
\end{figure}

We begin by examining how female and male users distribute across the detected communities (Figure~\ref{fig:gender-distributions-communities}).

By identifying the majority gender within each subreddit (considering only users belonging to the MBIA core), we find that 9 communities are predominantly female (left portion of the x-axis, from AskWomenOver30 to IncelTear), while 7 communities are predominantly male (from IncelTears to AskMenAdvice).
We emphasize that this classification pertains only to MBIA users active in these communities and does not necessarily reflect the overall gender composition of the subreddits.

\begin{figure}
\centering
\begin{subfigure}[b]{\columnwidth}
\includegraphics[width=\columnwidth]{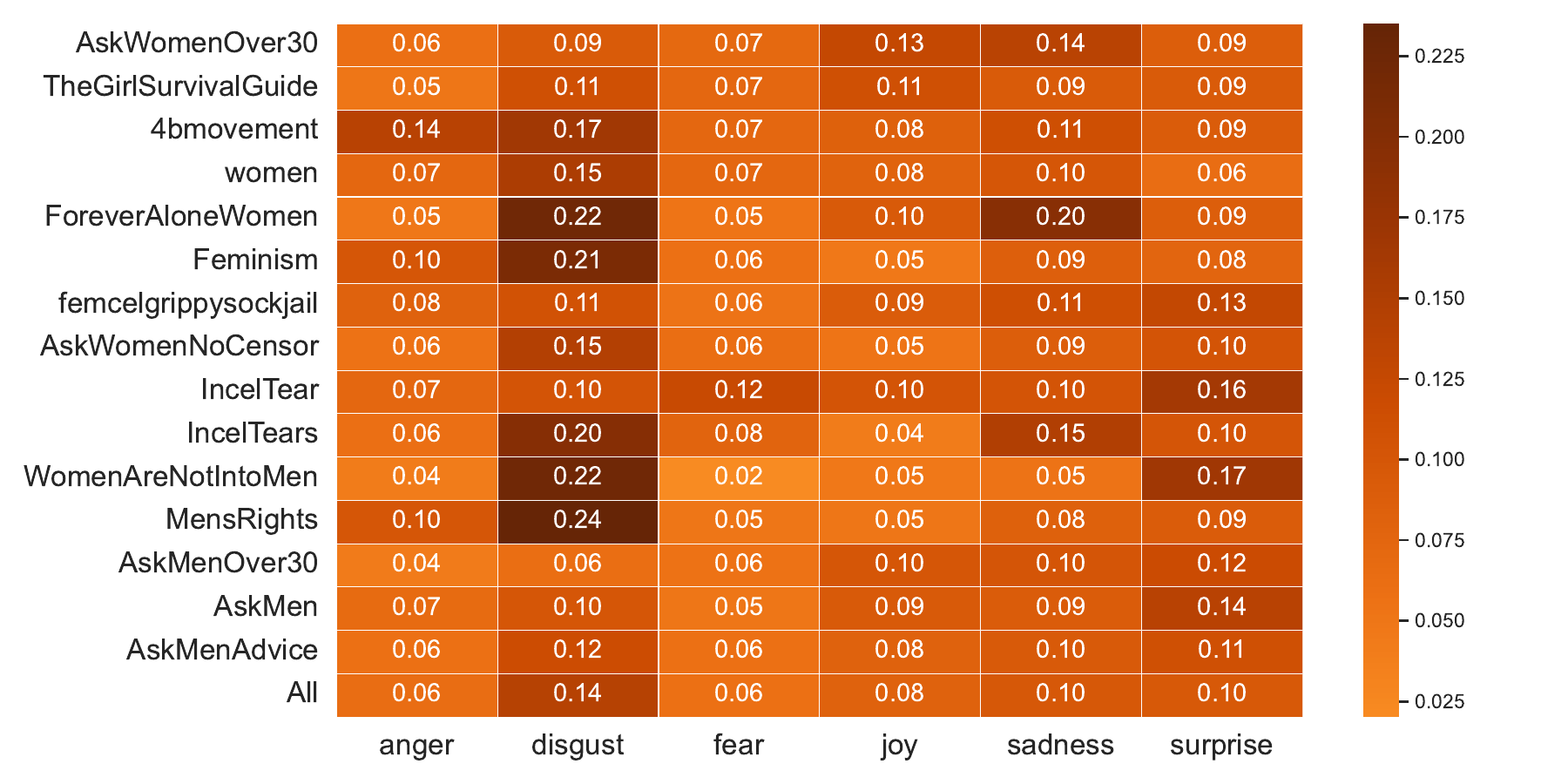}
\end{subfigure}
\begin{subfigure}[b]{\columnwidth}
\includegraphics[width=\columnwidth]{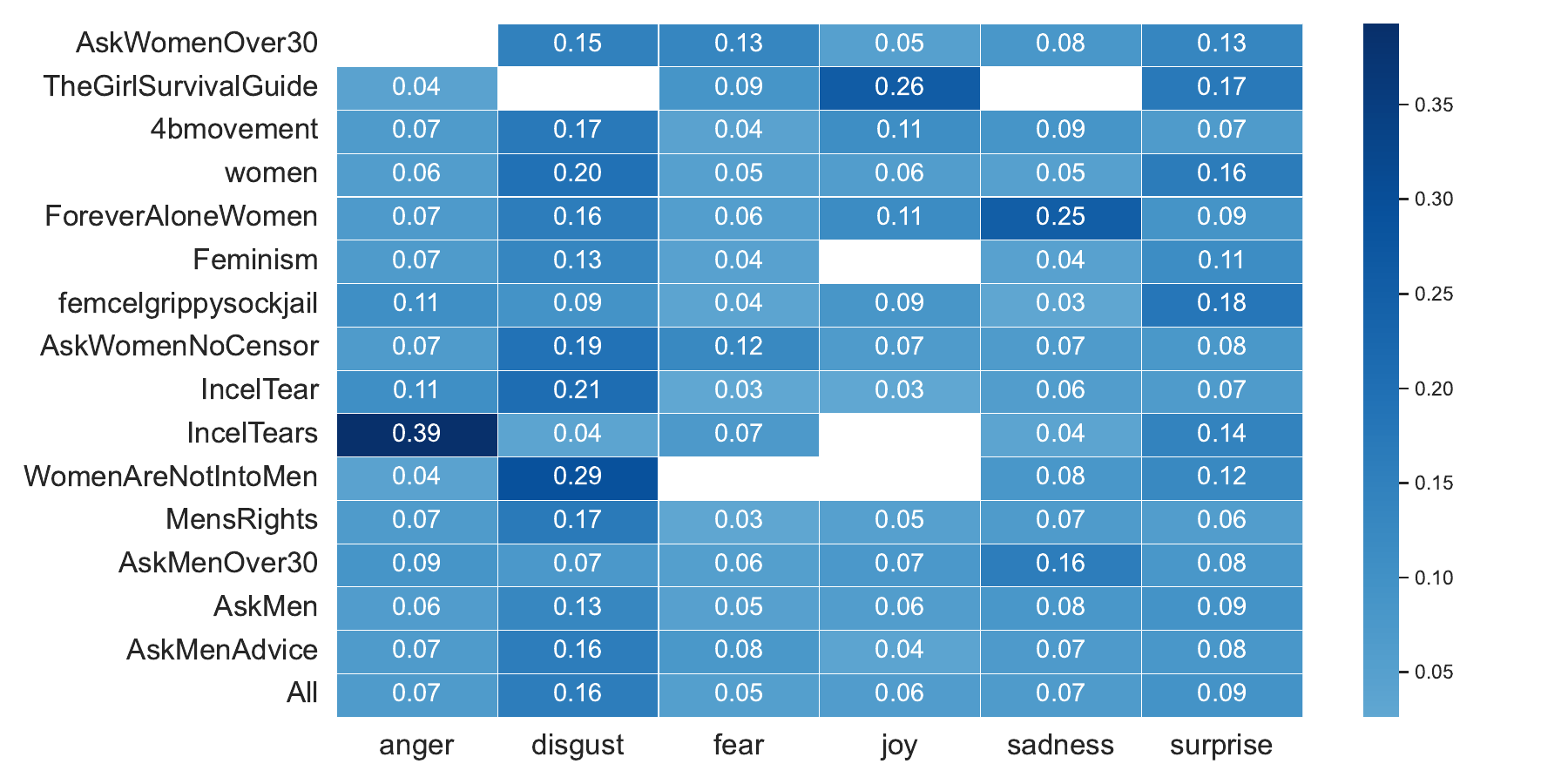}
\end{subfigure}
\caption{Proportion of female/male authors displaying the given emotion in gender-oriented subreddits. Blank cells indicate $0$-values.}
\label{fig:emotions_extremist_gendered_communities}
\end{figure}

Next, we compare the principal emotions expressed by female and male authors within these communities. We use a distilled RoBERTa-based model~\cite{hartmann2022emotionenglish} trained to classify the six Ekman basic emotions~\cite{ekman1999basic}—anger, disgust, fear, joy, sadness, and surprise—plus a neutral class. Figure~\ref{fig:emotions_extremist_gendered_communities} displays, for each gender-oriented subreddit, the proportion of female and male authors expressing each non-neutral emotion. The bottom row (“All”) shows the median proportion across all communities. Neutral predictions are omitted to highlight more meaningful affective patterns.

For female authors, the most prevalent emotion is disgust (median 12\%), especially pronounced in predominantly male communities such as MensRights, WomenAreNotIntoMen, and IncelTears, as well as in female-oriented subreddits such as ForeverAloneWomen and Feminism. The next most common emotions are surprise (11\%) and sadness (10\%).

For male authors, disgust also emerges as the dominant emotion (16\%), peaking in WomenAreNotIntoMen, followed by IncelTear and women. Other frequent emotions include surprise and fear (8\%), followed by anger and sadness (7\%).

Interestingly, male authors reach their highest anger levels specifically in the IncelTears community, whereas female authors exhibit a more consistent pattern of disgust across several subreddits.

\begin{figure}[!ht]
\centering
\includegraphics[width=\linewidth]{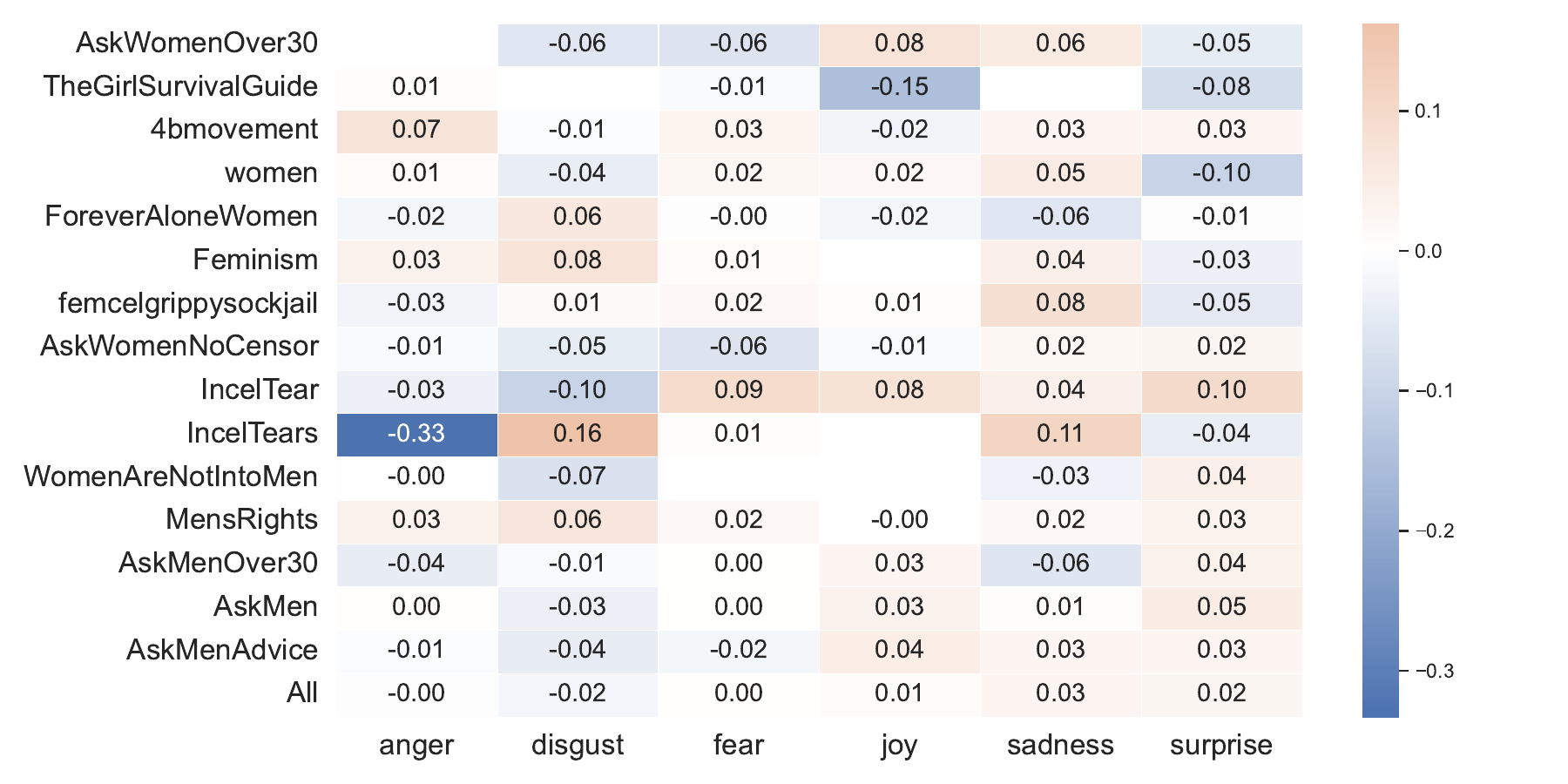}
\caption{Difference in proportion between female and male authors across gender-oriented subreddits. Positive (resp. negative) values indicate emotions more prevalent in female- (resp. male-) authored content. Neutral sentiment omitted.}
\label{fig:emotions-differnece}
\end{figure}

To directly compare the emotional expressions of women and men, Figure~\ref{fig:emotions-differnece} shows the difference in proportions between female and male authors for each emotion and subreddit. Positive values indicate emotions expressed more frequently by women; negative values correspond to emotions more frequently expressed by men.

Once again, IncelTears stands out as the most polarized community, showing the largest gender gaps: a 33\% excess of anger among male authors and a 16\% excess of disgust among female authors.
Overall, we observe that disgust is more frequently expressed by men than women, whereas women more often express sadness, surprise, and joy. A noteworthy cross-pattern emerges for surprise: men express it more often in predominantly female communities, while women express it more often in predominantly male ones.   

\begin{figure}[!ht]
    \centering
    \includegraphics[width=\linewidth]{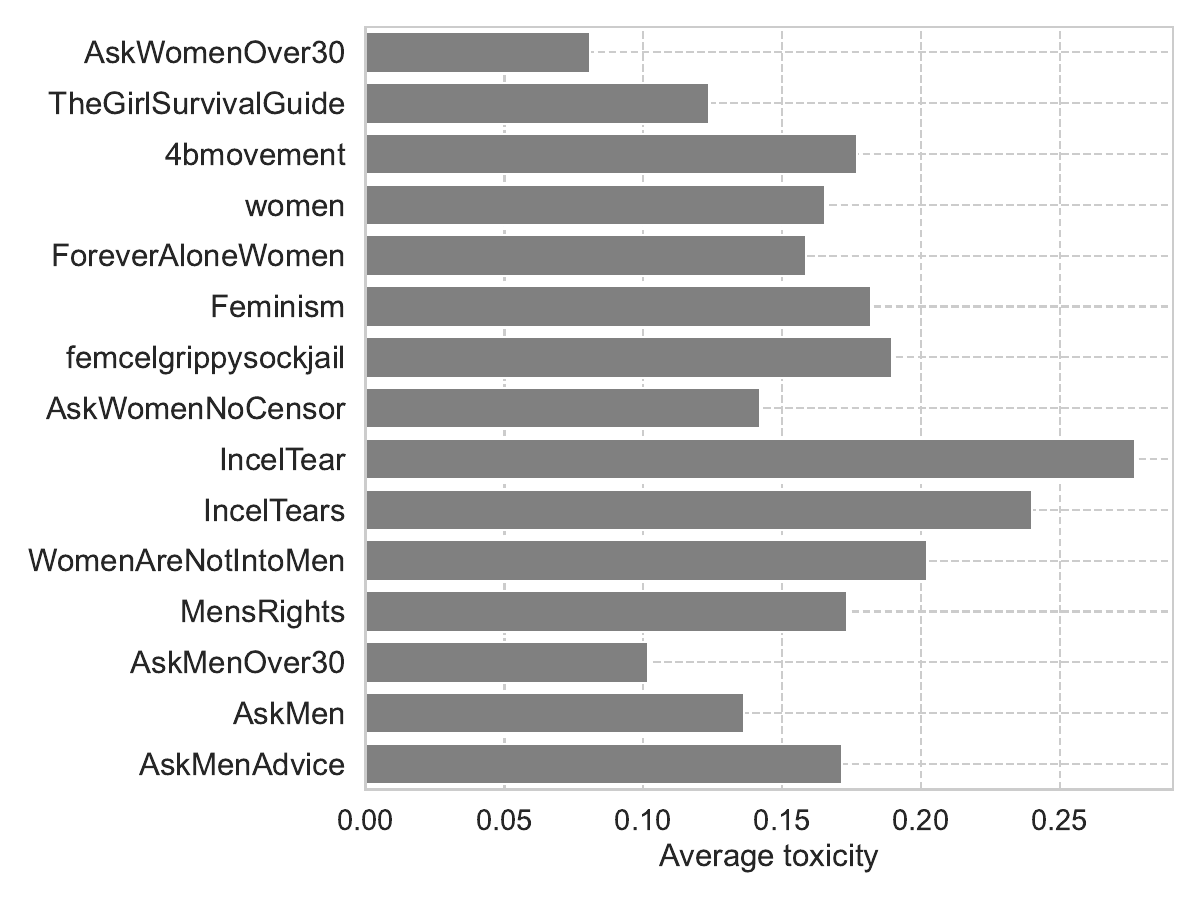}
    \caption{Average toxicity score of each gender-oriented subreddit.}
    \label{fig:toxicity-gendered-subreddits}
\end{figure}
Next, we examine the \textit{toxicity} associated with the gender-oriented subreddits. Figure~\ref{fig:toxicity-gendered-subreddits} reports the average toxicity score of each community, computed by averaging the toxicity of all posts and comments authored within that subreddit.

To assess whether toxicity varies by user gender, we further compute the toxicity gap between content written by female and male authors. Positive values indicate higher toxicity among female-authored content; negative values indicate higher toxicity among male-authored content.

Interestingly, most communities exhibit higher toxicity from male authors, including those that are predominantly female. Two exceptions emerge: ForeverAloneWomen (average gap $\sim$0.15) and 4bmovement ($\sim$0.14), where female-authored content is more toxic. These findings suggest that, on average, male MBIA users are more likely to produce toxic content, though meaningful exceptions appear depending on the subreddit context.

\begin{figure}[!ht]
\centering
\includegraphics[width=\linewidth]{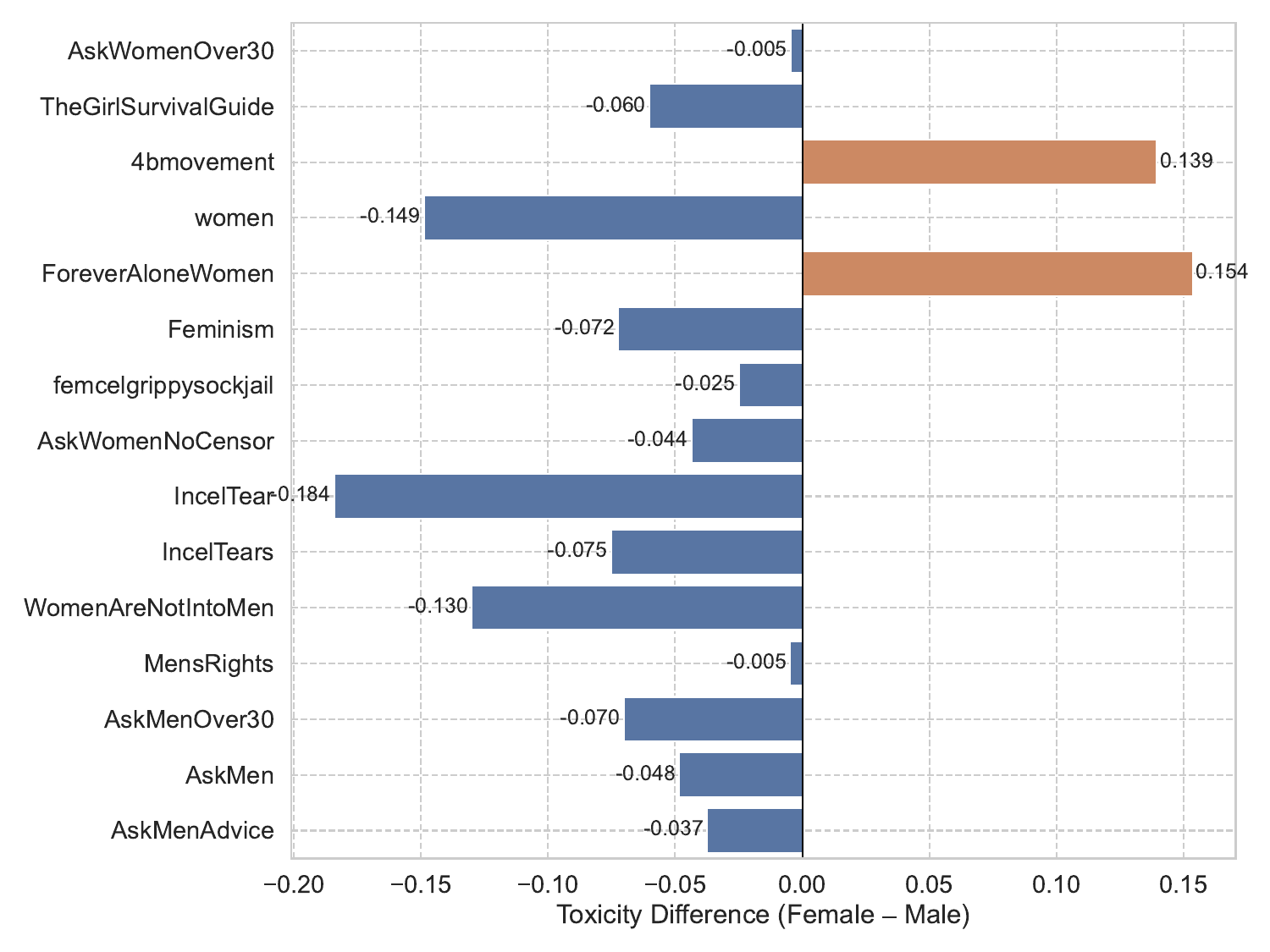}
\caption{Average toxicity gap between female- and male-authored content in gender-oriented subreddits. Positive (resp. negative) values indicate higher toxicity in female- (resp. male-) authored content.}
\label{fig:toxicity-gendered-subreddits-male-female}
\end{figure}

We further compare the toxicity levels of these users submissions across MBIA, gender-oriented subreddits, and their other historical communities. This analysis evaluates whether these users are generally toxic or whether their toxicity is context-dependent.

\begin{figure}[!ht]
\centering
\includegraphics[width=\linewidth]{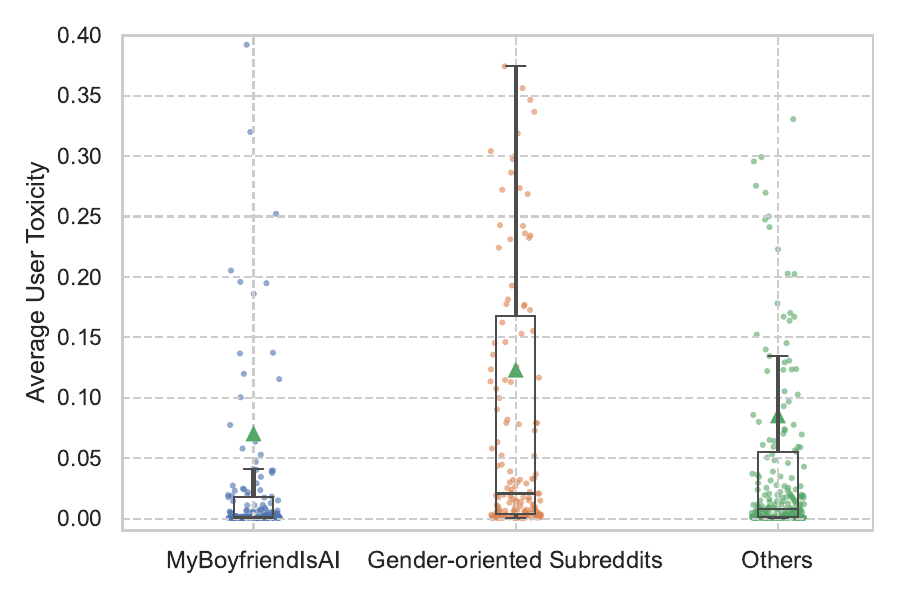}
\caption{Toxicity distribution across MBIA, gender-oriented subreddits, and other historical communities for MBIA users active in gender-oriented subreddits.}
\label{fig:toxicity-mbia-gendered-subreddits-others}
\end{figure}

Figure~\ref{fig:toxicity-mbia-gendered-subreddits-others} shows that toxicity is lowest on MBIA, higher on the users other historical subreddits, and highest in gender-oriented communities. This pattern indicates that the toxic behavior of these users cannot be detected by examining the MBIA subreddit alone, but only by considering their broader Reddit history.

To better contextualize these results, we compare the toxicity distribution of users active in gender-oriented subreddits with that of a random sample of MBIA users of equal size who never posted in gender-oriented communities.

\begin{figure}[!ht]
\centering
\includegraphics[width=\linewidth]{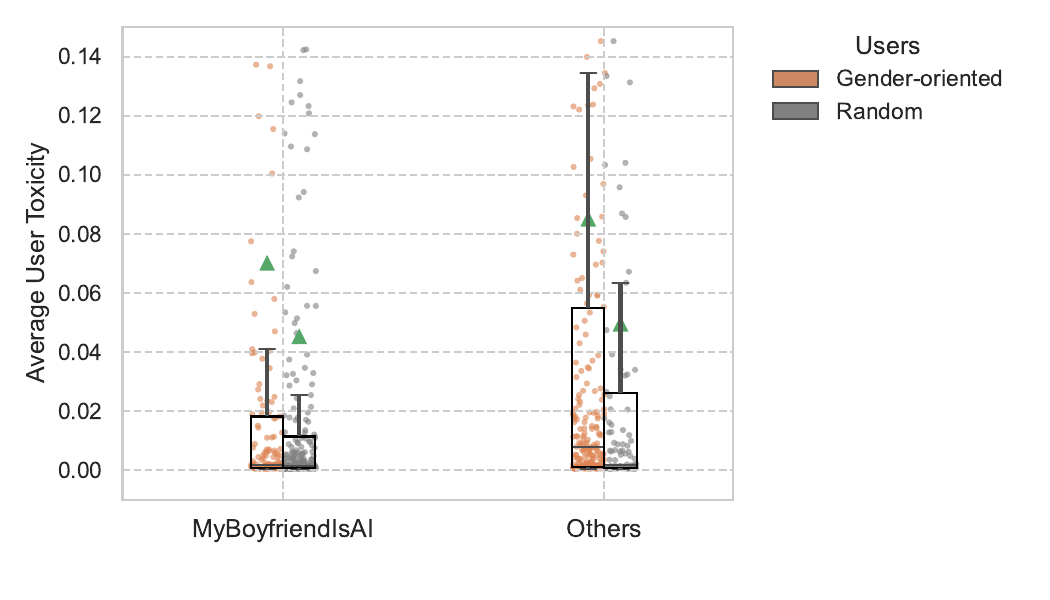}
\caption{Toxicity distribution for MBIA users active in gender-oriented subreddits and a random matched sample of MBIA users who are not.}
\label{fig:comparison-toxicity-gendered-random-userrs}
\end{figure}

Figure~\ref{fig:comparison-toxicity-gendered-random-userrs} shows that the toxicity distribution of users active in gender-oriented communities is consistently and significantly higher than that of the random sample, both on MBIA and on their other historical subreddits. This reinforces the conclusion that users active in gender-oriented communities exhibit systematically more toxic behavior.

Finally, we compare the community preferences of these two user groups. We compute the Jaccard similarity between the sets of subreddits where each group posts (excluding MBIA and the gender-oriented subreddits). The similarity is low (0.19), indicating distinct community engagement patterns. To further examine these differences, we compare the top-20 most popular subreddits for each group (Table~\ref{tab:placeholder}).

\begin{table}[!ht]
\centering
\caption{Top-20 most popular subreddits in the historical submissions of MBIA users who are/are not active in gender-oriented communities. For each subreddit, the percentage of users posting is reported.}
\label{tab:placeholder}
\resizebox{\columnwidth}{!}{
\begin{tabular}{@{}l@{}cl@{}c@{}}
\toprule
\multicolumn{2}{c}{Gender-oriented Users} & \multicolumn{2}{c}{Random Users} \\
\cmidrule(lr){1-2}
\cmidrule(lr){3-4}
Subreddit & Users Posting (\%) & Subreddit & Users Posting (\%) \\
\midrule
TrueUnpopularOpinion & 3.3 & SoraAi & 1.1\\
duolingo & 3.0 & indiehackers & 0.8\\
Christianity & 2.4 & ClashRoyale & 0.6\\
WitchesVsPatriarchy & 2.2 & Schedule\_I & 0.6\\
CleaningTips & 2.2 & OkBuddyFresca & 0.6\\
PurplePillDebate & 2.0 & HouseOfTheDragon & 0.6\\
piercing & 2.0 & lifehacks & 0.6\\
AskVet & 2.0 & Telegram & 0.6\\
aliens & 2.0 & citypop & 0.6\\
atheism & 1.7 & DMT & 0.6\\
recruitinghell & 1.7 & DatingInIndia & 0.6\\
ForeverAlone & 1.7 & betatests & 0.6\\
gamingsuggestions & 1.5 & kpophelp & 0.6\\
AuDHDWomen & 1.5 & gbstudio & 0.6\\
introvert & 1.5 & femalelivingspace & 0.6\\
AskConservatives & 1.5 & flashlight & 0.4\\
askatherapist & 1.5 & MVPLaunch & 0.4\\
NotHowGirlsWork & 1.5 & OpenAIDev & 0.4\\
suggestmeabook & 1.5 & ProductivityApps & 0.4\\
AiUncensored & 1.5 & ApplyingToCollege & 0.4\\
\bottomrule
\end{tabular}
}
\end{table}

The comparison reveals substantial differences between the two groups.
Users active in gender-oriented communities show a more concentrated and top-heavy distribution, engaging disproportionately with subreddits focused on gender debates (WitchesVsPatriarchy, PurplePillDebate, NotHowGirlsWork), religion/mysticism (Christianity, atheism, aliens), mental health and loneliness (ForeverAlone, AuDHDWomen, askatherapist), and political or contentious discussions (TrueUnpopularOpinion, AskConservatives).

In contrast, the random user group displays a more diverse and benign subreddit landscape, with interests centered on AI platforms (SoraAi, OpenAIDev), gaming and media (ClashRoyale, HouseOfTheDragon), and daily-life or hobby communities (lifehacks, DatingInIndia, femalelivingspace, ApplyingToCollege).

\section{Discussion}

Our findings show that AI companionship does not emerge in isolation but is embedded in a diverse ecosystem encompassing AI-related communities, porn-oriented content, emotional support groups, and gaming forums. This demonstrates the need for ecosystem-level analysis of AI-mediated intimacy rather than a platform-specific perspective.

The strong predominance of female users in MBIA (including in porn-oriented spaces) challenges popular assumptions that AI romance primarily attracts male or incel-associated populations. Differences in emotional expression between male and female authors further suggest that AI companionship intersects with broader gendered online behaviors and vulnerabilities.
While overall toxicity in user trajectories is low, a meaningful subset of MBIA users participates in toxic or gender-hostile communities. These users show significantly higher toxicity across MBIA, gender-oriented communities, and their remaining historical activity, suggesting that gender-oriented spaces might operate as toxicity amplifiers.

We believe our work raises important implications for AI safety and platform design. Indeed, AI companionship systems may unintentionally reinforce harmful emotional patterns or validate toxic beliefs through conversational mirroring. Platform designers and moderators may benefit from understanding cross-platform trajectories, identifying at-risk users, and designing interventions to prevent harmful reinforcement loops.

\section{Conclusions}

In the present work, we reconstruct the multi-year Reddit histories of more than 3,000 core users of the \textit{MyBoyfriendIsAI} subreddit, generating the first ecosystem-level map of user pathways into AI-romantic communities. Our findings reveal that MBIA users navigate through four dominant communities (AI companionship, porn-oriented content, forum discussions, and gaming) and that the user base is unexpectedly female-dominated.  
Toxicity is generally low, but spikes within a subset of gender-oriented and AI-porn communities, and emotional expression varies substantially across genders.  
These findings highlight that AI romantic companionship is intertwined with a broader, sometimes contentious ecosystem of online participation. Our work provides an empirical foundation for understanding emergent risks and opportunities in human-AI relationships, offering implications for moderation, AI safety, and the design of emotionally interactive agents.

\paragraph{Limitations} Despite providing some insights, the analysis presents the following limitations. First, we perform toxicity and gender inference by relying on automatic models that, although accurate, might introduce biases. We also emphasize that the gender classifier estimates a \textit{linguistic} gender presentation, not identity. Further, Reddit data excludes private interactions or off-platform activity, preventing us to generalize our findings beyond the Reddit ecosystem.

\paragraph{Future Work} The present work is amenable to further improvements. One promising direction is to investigate how users participation in gender-oriented subreddits shapes, mediates, or moderates their relationship with AI companions, potentially influencing patterns of attachment, self-disclosure, or identity expression. Complementarily, future studies could adopt qualitative or mixed-methods approaches to gain deeper insight into users subjective experiences and motivations that are not fully captured by quantitative analysis alone. Expanding the scope of analysis to include multi-platform datasets would also allow researchers to assess whether observed behaviors generalize across different social media environments or are platform-specific. Finally, longitudinal and causal inference methods could be employed to evaluate the extent to which AI companionship influences subsequent human behavior on social platforms, shedding light on the broader social and psychological implications of sustained human-AI interaction.

\section*{Acknowledgments}
% Anonymized
This work has been partially funded by MUR on D.M.\ 351/2022, PNRR Ricerca, CUP H23C22000440007, further in part supported by the NSF (Award Number 2331722).

% Apparently, the package is forbidden
% \balance
\bibliography{ref}

\appendix

\end{document}